\documentclass[twocolumn]{aastex631}
\usepackage{stfloats}
\usepackage{float}
\usepackage{graphicx}
\usepackage{natbib}
\usepackage{hyperref}
\usepackage{amsmath}
\usepackage{autobreak}
\usepackage{multirow}
\usepackage{makecell}
\usepackage{color}
\usepackage{bm}
\usepackage{appendix}
\usepackage{threeparttable}
\usepackage[figuresright]{rotating}
\usepackage[mathscr]{euscript}
\received{}
\revised{}
\accepted{May 3rd, 2021}
\submitjournal{ApJ}

\shorttitle{}
\shortauthors{Niu et al.}

\graphicspath{{./}{figures/}}
\begin{document}

\title{Not that simple: the metallicity dependence of the wide binary fraction changes with separation and stellar mass}

\correspondingauthor{Zexi Niu}
\email{nzx@nao.cas.cn}
\correspondingauthor{Haibo Yuan}
\email{yuanhb@bnu.edu.cn}

\author[0000-0002-3651-0681]{Zexi Niu}
\affiliation{National Astronomical Observatories,
Chinese Academy of Sciences, 20A Datun Road, Chaoyang District, Beijing, China}
\affiliation{University of Chinese Academy of Sciences, Yuquan Road, Shijingshan District, Beijing, China}

\author[0000-0003-2471-2363]{Haibo Yuan}
\affiliation{Department of Astronomy,
Beijing Normal University, 19th Xinjiekouwai Street, Haidian District, Beijing, China}

\author[0000-0002-3530-3277]{Yilun Wang}
\affiliation{National Astronomical Observatories,
Chinese Academy of Sciences, 20A Datun Road, Chaoyang District, Beijing, China}
\affiliation{University of Chinese Academy of Sciences, Yuquan Road, Shijingshan District, Beijing, China}

\author{Jifeng Liu}
\affiliation{National Astronomical Observatories,
Chinese Academy of Sciences, 20A Datun Road, Chaoyang District, Beijing, China}
\affiliation{University of Chinese Academy of Sciences, Yuquan Road, Shijingshan District, Beijing, China}

\begin{abstract}

The metallicity dependence of the wide binary fraction (WBF) is critical for studying the formation of wide binaries.  
While controversial results have been found in recent years.
Here we combine the wide binary catalog recognized from \textit{Gaia} EDR3 and stellar parameters from LAMOST to investigate this topic. 
Taking bias of the stellar temperature at given separations into account, we find that the relationship between the WBF and metallicity depends on temperature for the thin-disk at $s > 200$\,AU. It changes from negative to positive as the temperature increases from 4000\,K to 7500\,K. 
This temperature/mass dependence is not seen for the thick-disk. 
Besides, the general tendency between the WBF and metallicity varies with the separation,
consistent with previous results. 
It shows anti-correlation at small separations, $s < 200$\,AU for the thin-disk and $s < 600$\,AU for the thick-disk. Then it becomes an ``arcuate'' shape at larger separations (hundreds to thousands of AU), peaking at [Fe/H] $\approx$ 0.1 for the thin-disk and [Fe/H] $\approx-0.5$ for the thick disk. Finally it becomes roughly flat for the thin-disk at $1000<s<10000$\,AU. 
Our work provides new observational evidences for theoretical studies on binary formation and evolution. 

\end{abstract}

% \keywords{}

\section{Introduction} \label{sec:intro}

Binary systems are ubiquitous in our universe. According to their separations, they are empirically classified into two groups: close binaries and wide binaries. In most cases, wide binaries refer to spatially resolved and comoving pairs, whose separations depend on their distances to us, ranging from 10$^1$\,AU to 10$^5$\,AU. 
The two components of a wide binary are thought to be born at the same time and place \citep{1993goodman,2004Fisher,2010Kouwenhoven}. Therefore, they are believed to have the same age and metallicity in principle.
Their evolution could be treated as isolated single stars due to the large physical separation, it makes them as ideal laboratories for testing stellar models and calibrating gyrochronology relations \citep{Gyrochronology}. 
As the binding energy of wide binaries is relatively small, they are sensitive to the gravitational perturbations in the Milky Way, making them good indicators for the study of Galactic dynamics \citep{1975Heggie,macho,2010jiang,2014Kaib}.

Identifications of wide binaries are challenging. Previous samples of wide binaries are small, and mostly limited to the solar neighborhood ($d\le100$ pc) with separations less than 10$^3$\,AU (e.g, \citealp{2007ls,raghavan2010,2014TK}).
Thanks to the whole sky survey \textit{Gaia} \citep{gaia2016}, a large number of wide binaries can be identified through the common proper motion technique (e.g,  \citealp{2020Hartman,2017pw,2019jm}). 
Using data from \textit{Gaia} EDR3 \citep{gaiaEDR3}, \citet{2021plx} constructed the largest catalog of spatially resolved binary stars 
within 1 kpc of the Sun. 
These much larger samples enable studies of binary properties in great detail, such as the distributions of binary separation \citep{2017Andrews,2018ekwide,2020tian}, mass-ratio \citep{2019twin}, and eccentricity \citep{EDR3edis}.

The detailed metallicity dependence of the binary fraction plays a fundamental role in constraining binary formation mechanisms. 
Unlike the well-established anti-correlation between the (close) binary fraction and metallicity (e.g, \citealp{raghavan2010,2015Yuan,gao2017,2018badnes,2018Tian,moe2019anti,niu_bf}), some recent works have shown that the metallicity dependence of the wide binary fraction (WBF) may be different.
Using comoving pairs within 200 pc from \textit{Gaia} DR2 \citep{2018dr2wide} and metallicities from 5 spectroscopic surveys (LAMOST, RAVE, APOGEE, GALAH, and Hypatia),
\citet{widedr2} found that even though the WBF displays an anti-correlated behavior with [Fe/H] at small separations ($s < 200$) which is similar to the close binary fraction, it becomes generally constant with [Fe/H] at larger separations. 
\citet{widefb2021} revisited the metallicity dependence of field wide binaries ($s = 10^3-10^4$\,AU) using LAMOST and \textit{Gaia} DR2 data sets. They limited sample to 5000 -- 7000\,K and extended the sample out to 500 pc. They also grouped the sample into the thin-disk, the thick-disk, and the halo stars according to the kinematic definition.
An entirely different conclusion was drawn: as metallicity increases, the WBF increases firstly but then decreases at the high-metallicity end. They demonstrated that the tendency steepens in the thin-disk than in the thick-disk, the WBF of the thin-disk peaks around the solar abundance while the WBF of the thick-disk peaks  --  $-$0.5.
Taking the kinematics of stars as age proxy, they also showed that younger stars tend to have a higher WBF around solar abundance.

It's reasonable to speculate that deficiencies in the previous works caused the discrepancies.
It could be better to take stellar metallicities from one spectroscopic survey thinking of possible offsets of stellar parameters among different surveys \citep{2017Nandakumar,2018Anguiano}. 
Both \citet{2018dr2wide} and \citet{widefb2021} used ``control samples" to calculate the WBF in their methods.  
\citet{2018dr2wide} constructed a distance-dependence control sample to avoid selection bias caused by unresolved close binaries. 
\citet{widefb2021} used the whole LAMOST stars as the their control sample, ignoring the influences from the distance-dependent unresolved close binaries. 
What's more, the trend that the total binary fraction increases with the stellar mass has been widely confirmed (\citealp{2013araa} and references therein). The same trend is also found for wide binaries \citep{2017Moe,2021gaianb}.
Unfortunately, bias of the stellar temperature at given binary separations is ignored by them. Besides, neither of them considered systematic errors of the spectroscopic metallicities.

In this paper, 
we combine the wide binary catalog recognized from \textit{Gaia} EDR3 and stellar parameters from LAMOST to investigate the metallicity dependence of the WBF. 
Section \ref{secdata} describes data selection and processing. 
Here we use revised metallicities.
Section \ref{secmr} demonstrates method and results. 
``Control samples" are carefully constructed to identify the WBF tendency with metallicity. 
Section \ref{secdis} discusses caveats in this work and compares with previous works.
Section \ref{summary} summarizes the paper.

\section{Data}\label{secdata}

The wide binary catalog published by \citet{2021plx} contains over 1 million binaries within 1 kpc and with projected separations from a few AU to 1 pc\footnote{The median-projected separation $s$ is approximately equal to the semi-major axis $a$ \citep{2010jiang,2021plx}.}. 
We start with main-sequence/main-sequence pairs and required their rate of chance alignments $ \mathscr{R} < 0.1$ to select high-confidence ones.
The selected wide binaries are then cross-matched with the value-added catalog of LAMOST DR5 of \citet{xiangdd} to 
obtain their chemical abundances.
We further require signal-to-noise ratio of spectra larger than 20. 
% A total number of 78,152 binary systems are obtained.

\citet{xiangdd} used a data-driven Payne approach (DD-Payne) to deliver stellar parameters as well as abundances for 16 elements of 6 million stars, training models from high-resolution surveys like GALAH and APOGEE. 
The precision of T$_{\rm eff}$ and [Fe/H] are about 50\,K and 0.06, respectively.
[$\alpha$/Fe] is defined as a weighted mean of [Mg/Fe], [Si/Fe], [Ca/Fe], and [Ti/Fe] with precision of about 0.03.

We have preformed a diagnosis of chemical abundances of several spectroscopic surveys in Niu et al. (in prep) using wide binaries. 
According to the formation theories of wide binaries (e.g. \citealp{2004Fisher}), components of a wide binary are formed at the same time and place, and have the same initial surface abundances.
For F/G/K type stars, influence of atomic diffusion during the main-sequence stage on the surface abundances (e.g. [Fe/H]) is only about 0.05 dex on average \citep{2017Dotter,2022Campilho}. 
This effect is much smaller than the possible systematic errors in spectroscopic surveys.
Therefore, wide binaries in which both components are targeted can be used to examine chemical abundances of spectroscopic surveys.
We selected main-sequence binaries of F/G/K type with spectral signal-to-noise ratio larger than 20. 
Here we briefly demonstrate the results of \citet{xiangdd}'s catalog.
Taking [Fe/H] as an example, we obtained empirical formulas of revised [Fe/H] by minimizing the loss function: $|$[Fe/H]$_{\rm rev,pri}-$[Fe/H]$_{\rm rev,sec}|$. 
Regularization was applied during the fitting to avoid over-corrections. 
The correction at 5770\,K was set to be zero.
We have validated our corrections using open clusters, whose member stars share the same surface abundances.

\begin{figure}[htbp] 
    \centering %使图片居中显示
    \includegraphics[width=3in]{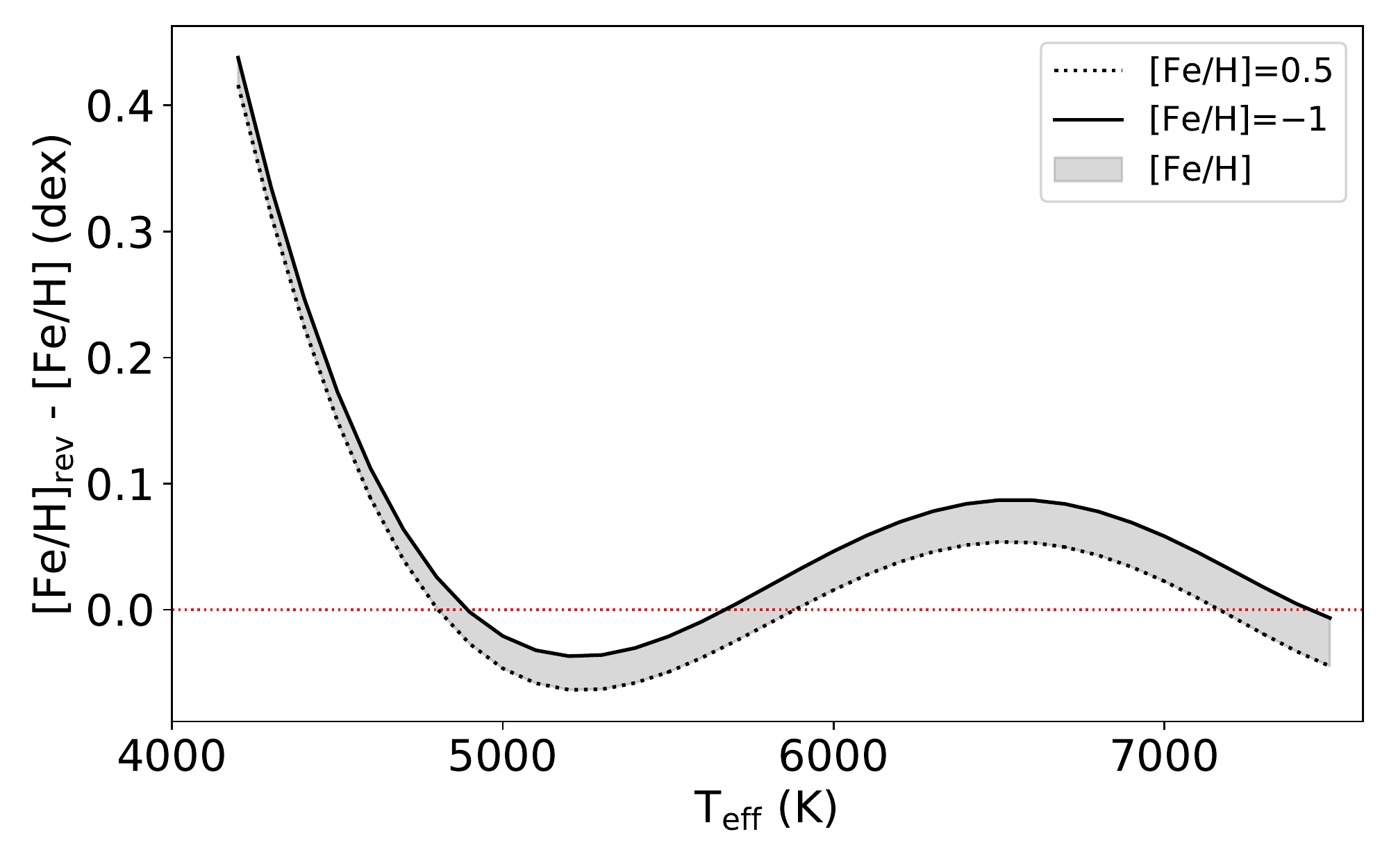}
    \caption{Variations of [Fe/H]$_{\rm rev}$ $-$ [Fe/H] against temperature. 
    Cases for published [Fe/H] equals to 0.5 and $-1$ are respectively plotted by solid and dotted lines. 
    Horizontal red line at zero is plotted for reference.
     \label{fehcorr}} 
\end{figure}

Coefficients for [Fe/H] and [$\alpha$/Fe] are listed in Table \ref{tab1} and plotted in Figure \ref{fehcorr}.
We found that the [Fe/H] and [$\alpha$/Fe] values are strongly affected by systematic errors. 
The errors are mostly temperature-dependent and partly depend on chemical abundances. 
There are significant under-estimations of [Fe/H] and over-estimations of [$\alpha$/Fe] for stars colder than 5000\,K, reaching $-$0.4\,dex for [Fe/H] and $+$0.08\,dex for [$\alpha$/Fe] at 4200\,K.
Significant over-estimations of [$\alpha$/Fe] are also for stars hotter than 7000\,K.
Bias of [Fe/H] can be hardly seen for stars hotter than 5000\,K. 
We can imagine that this calibration is vital for stars colder than 5000\,K. 
This is confirmed in Section \ref{subseccomp}.
Corrections should be used with cautions for stars outside the ranges in Figure \ref{fehcorr}.
All [Fe/H] and [$\alpha$/Fe] values mentioned hereafter are revised values.

\begin{figure}[htbp] 
    \centering %使图片居中显示
    \includegraphics[width=3.5in]{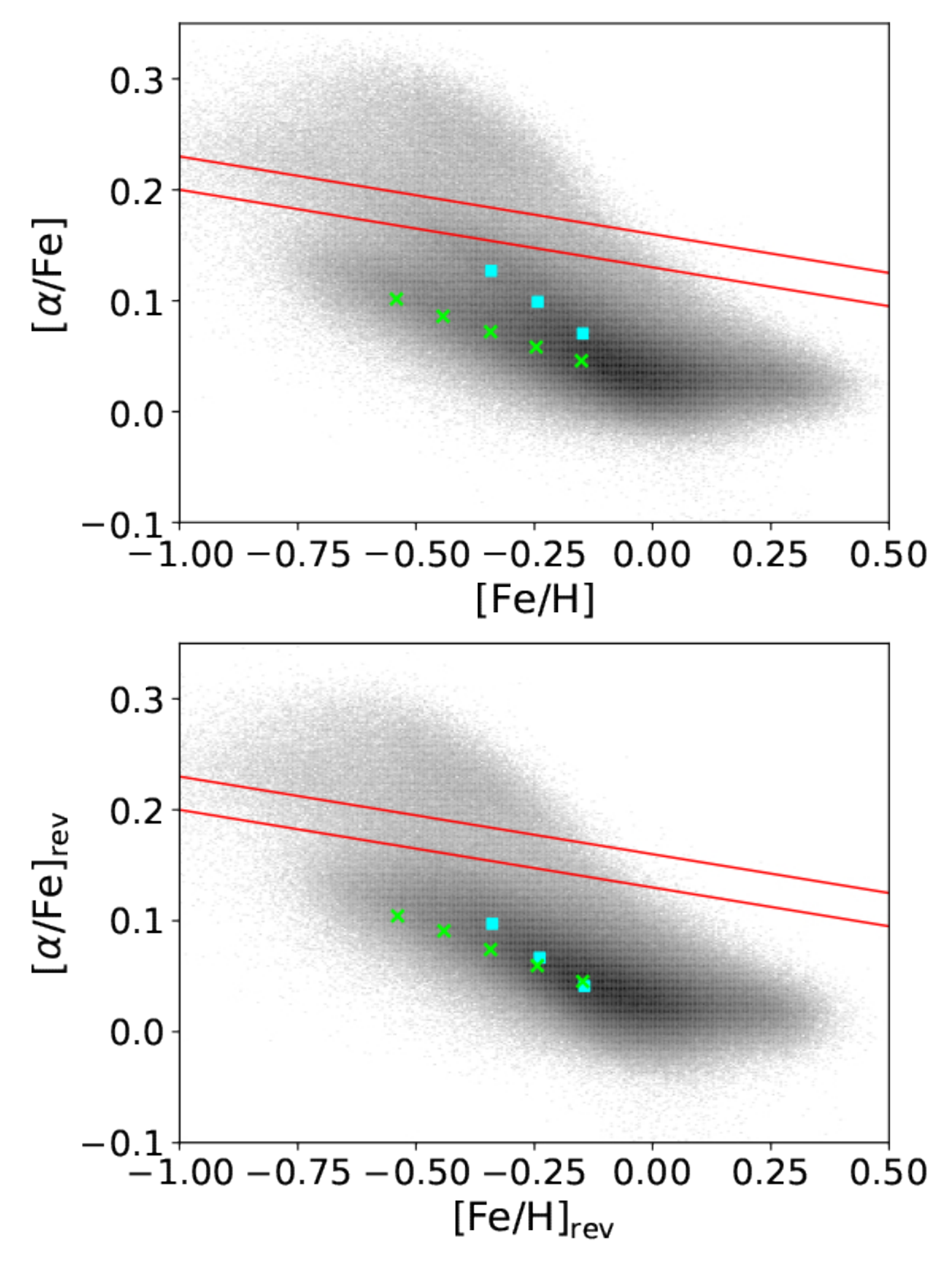}
    \caption{Density distributions in the [Fe/H]--[$\alpha$/Fe] plane of stars with high-quality data from the DD-Payne catalog. 
    Squares and crosses are median values of stars with 4500\,K  --  5000\,K and 6500\,K  --  7000\,K respectively. 
    Top panel is plotted with the published abundances, bottom panel is plotted with the revised abundances. Red lines are empirical cuts of the thin and thick disks. 
    \label{fehafe}} 
\end{figure}

Improvements of revised chemical abundances are demonstrated in Figure \ref{fehafe}. We select stars from the DD-Payne catalog with high-quality and plot their density distributions in the [Fe/H]--[$\alpha$/Fe] plane. 
Before the revision, high density area splits into two branches. Distributions of stars with 4500\,K  --  5000\,K are marked by cyan squares, whose [Fe/H] values are under-estimated and [$\alpha$/Fe] values are over-estimated. Therefore, they locate above stars with 6500\,K  --  7000\,K, shown by green crosses, which are less affected. 
When plotting with the revised abundances, the two branches merge into one, as expected.

For multiply observed \textit{Gaia} objects, we combine their epoch values using inverse variance weighting.
The mean parallax of two components is adopted as the parallax of the system. 
For most binary systems, only one component is spectroscopically matched.
According to the target selection process of LAMOST \citep{2015yuan_lasel}, they are probably the primary star of the systems. We have verified it using \textit{Gaia} colors. 
For systems where both components are spectroscopically matched, mean values of [Fe/H] are adopted, hotter stars are assigned to be the primary ones. 
We chemically divide the samples into thin-disk and thick-disk populations according to their locations in the [Fe/H]--[$\alpha$/Fe] plane. 
The following scheme are applied: 

Thick-disk: 
\begin{equation}
[\alpha /{\rm Fe}] > -0.05 \times [{\rm Fe/H}]+0.16 \nonumber
\end{equation}
Thin-disk:
\begin{equation}
[\alpha /{\rm Fe}] < -0.05 \times [{\rm Fe/H}]+0.13  \nonumber
\end{equation}

Verification of the disk populations is provided in Appendix.
Finally, we get the thin-disk sample containing 63,655 wide binaries, 2,489 of them having both spectroscopically-matched components. As for the thick-disk sample, numbers decrease to 3,347 and 86.

\begin{table*} [htbp]
    \centering
    \footnotesize
    \caption{Coefficients used to revise [Fe/H] and [$\alpha$/Fe] \label{tab1}}
    \begin{tabular}{c|c|c|c|c|c|c|c|c|c|c}
Coeff.$^1$ & $a_0$ & $a_1$ & $a_2$ & $a_3$ & $a_4$  & $a_5$ & $a_6$ & $a_7$ & $a_8$ & $a_9$  \\
\hline
[Fe/H] & 17.30 & $-$1.302 & $-$3.654  & 1.219  & $-$1.451e$-1$  &  6.011e$-3$ & - & $-$4.138e$-4$ & $-$2.729e$-3$ & 1.227e$-3$ \\
\hline
[$\alpha$/Fe] & $-$3.060 & 1.550 &  $-$2.300e$-1$ & $-$4.000e$-3$ & 3.500e$-3$ & $-$1.900e$-4$  & $-$3.326e$-$06 & - & - & - 
    \end{tabular}
    \begin{tablenotes}
    \item[1]$^1$$y_{\rm rev}=y+f(x,y)=y+a_6\times x^6+a_5\times x^5+a_4\times x^4+a_3\times x^3+a_2\times x^2+a_1\times x+ a_0+a_7\times y +a_8\times x \times y + a_9 \times x \times y^2$, where $x$ is T$_{\rm eff}/1000$ and $y$ is [Fe/H] or [$\alpha$/Fe] from the DD-Payne catalog.
    \end{tablenotes}
\end{table*}

\section{Method and result}\label{secmr}

\begin{figure}[H] 
    \centering %使图片居中显示
    \includegraphics[width=3.3in]{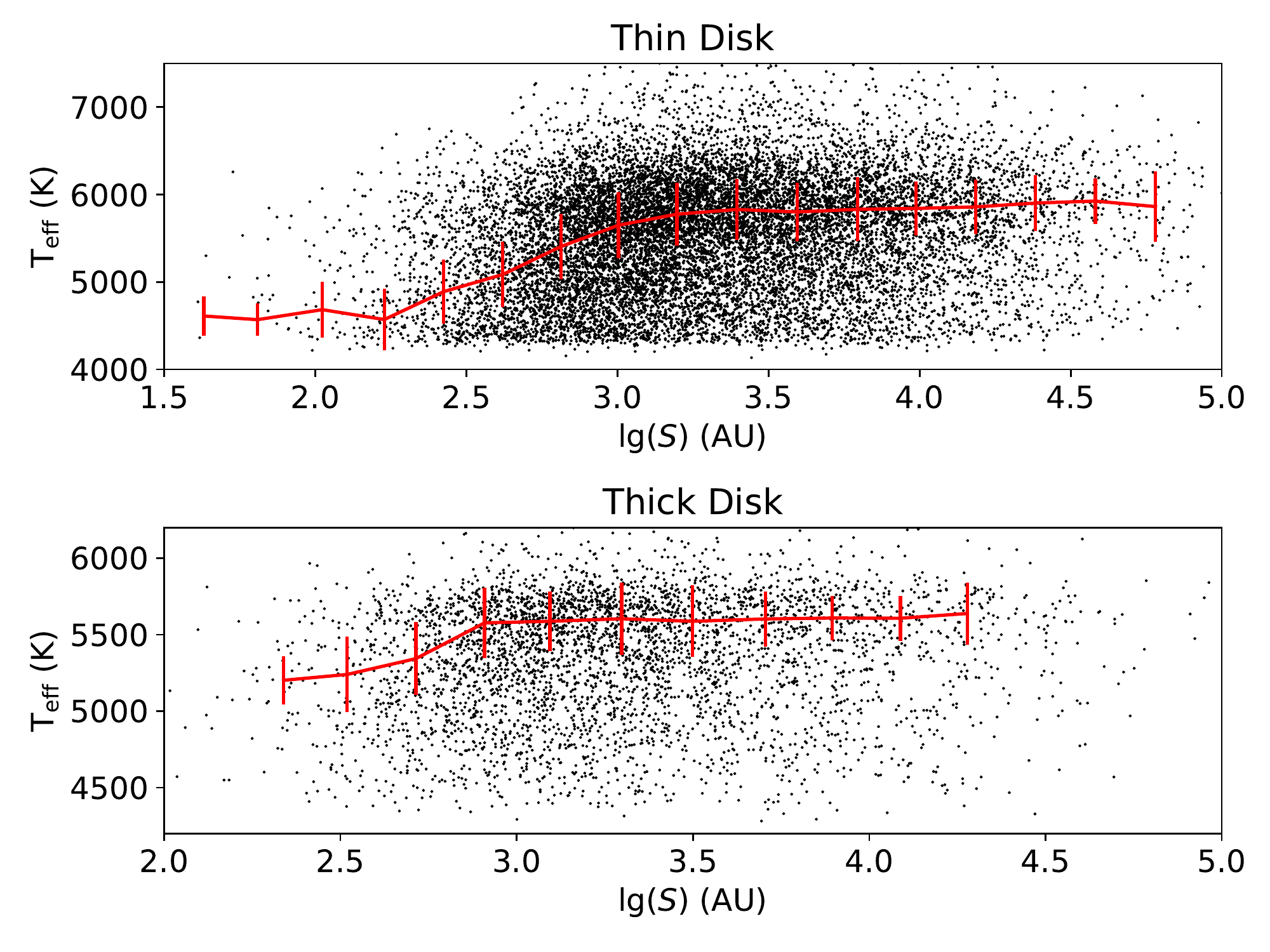}
    \caption{Distributions of the binary separation and the temperature of the primary of the thin-disk (top) and thick-disk (bottom) samples. 
    Medians and stranded deviations are plotted in red.
    Horizontal axes are in logarithmic scale.
    In the top panel, only 1 in 5 binaries are plotted.
    \label{teffsep}} 
\end{figure}

 \citet{widedr2} suggested that the relationship between WBF and [Fe/H] changes with the binary separation.
As shown in Figure \ref{teffsep}, distribution of the temperature of the primary also changes with the binary separation, especially for separations of hundreds of AU. 
There are more cold binaries in the samples with small separations than in samples with large separations, mainly due to the angular resolution of \textit{Gaia}. 
Therefore, we divide our samples into various subsamples according to the separations (0 -- 200\,AU, 200 -- 600\,AU, 400 -- 800\,AU, 600 -- 1000\,AU, and 1000 -- 10000\,AU) as well as temperatures (4000 -- 5000\,K, 4500 -- 5500\,K, 5000 -- 6000\,K, 5500 -- 6500\,K, 6000 -- 7000\,K, and 6500 -- 7500\,K).

In order to investigate the tendency of WBF on [Fe/H] of different subsamples, we compare the distribution functions of [Fe/H] of wide binaries to those of control samples, as described by \citet{widedr2}.  
We divide each subsample into ten bins with an equal interval of [Fe/H] ranging from $-1.0$ to $+$0.5.
Control samples are constructed by selecting common stars of \textit{Gaia} EDR3 and LAMOST DR5. They passed the same cuts that applied to the wide binary samples \citep{2021plx}. 
For each subsample of the wide binaries, its control sample is generated following the same distance and temperature distributions to avoid selection bias caused by unresolved binaries and physically different WBF of different stellar types. 
Then we calculate probability densities of [Fe/H] for the binary subsample and the control sample, $P_{\rm wide}$ and $P_{\rm control}$. 
The ratio of $P_{\rm wide}$ and $P_{\rm control}$ indicates the tendency of the WBF. 
For example, $P_{\rm wide,i}/P_{\rm control,i}=1$ means that the wide binaries and control samples have the same percentage at the $i_{\rm th}$ bin in [Fe/H]. $P_{\rm wide,i}/P_{\rm control,i} > 1$ means the percentage of the wide binaries increases but the percentage of the control samples decreases, i.e., a larger WBF.
To evaluate formal errors of $P_{\rm control}$, the Bootstrap Sample method is taken with 
2000 times of random sampling of control samples. Uncertainties of the $P_{\rm wide}/P_{\rm control}$ also include Poisson uncertainties of the $P_{\rm wide}$ and $P_{\rm control}$.
Notice that $P_{\rm wide}/P_{\rm control}$ represents the tendency of WBF instead of the absolute value.

\begin{figure*}[htbp] 
    \centering %使图片居中显示
    \includegraphics[width=7in]{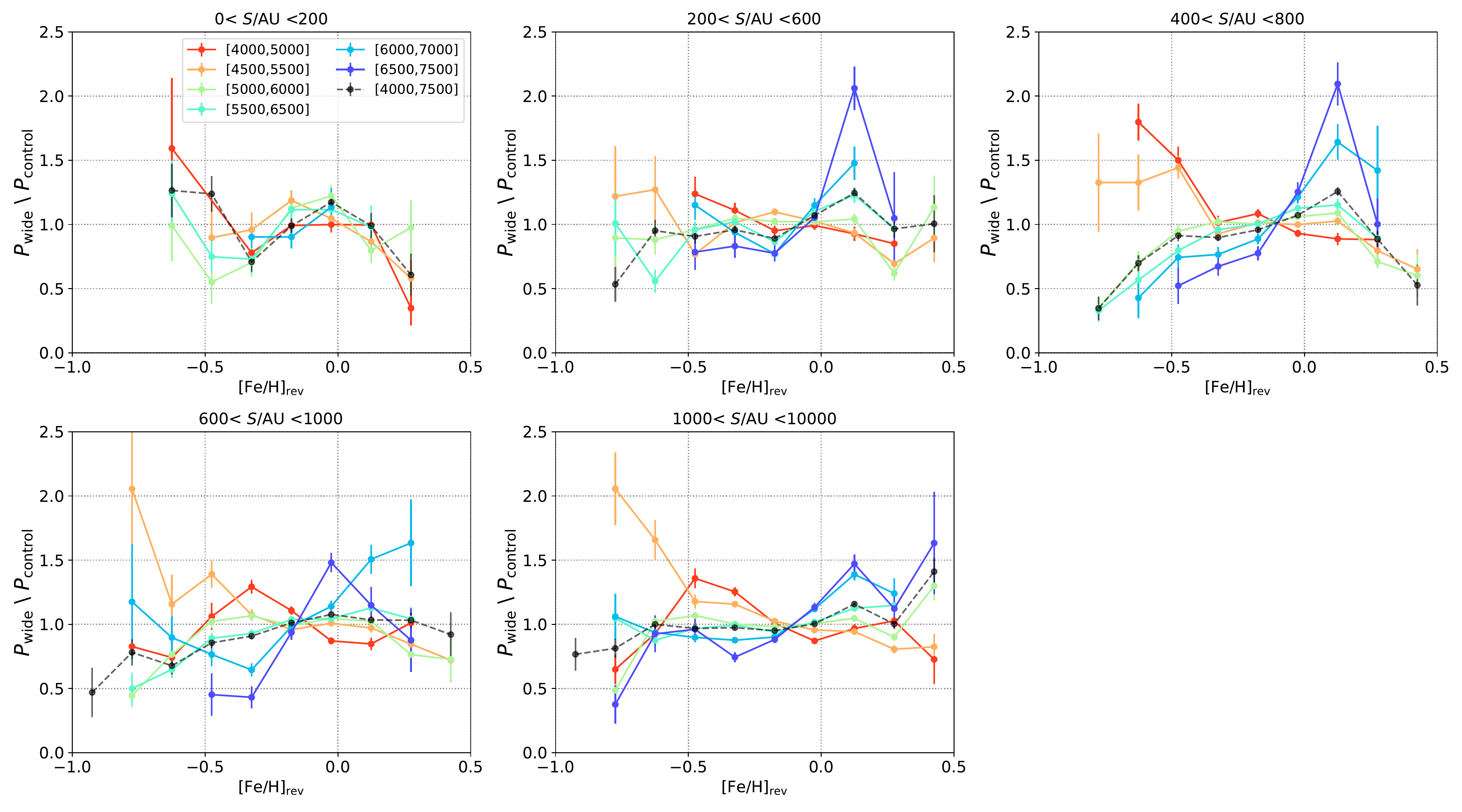}
    \caption{The WBF dependencies on [Fe/H] of the thin-disk sample with various temperatures of the primary and separations of the binary.
    $P_{\rm wide/control}$ are the probability density functions of the wide binaries/control samples. 
    Each control sample is randomly selected 2,000 times following the same distributions of distance and temperature as the wide binary. 
    Five panels correspond to five separation ranges, from 0 -- 200\,AU to 1000 -- 10000\, AU. 
    Colors indicate temperatures of the primary. 
    The general result of full temperature ranges are plotted in black. Vertical error bars include Poisson uncertainties and random errors obtained from the random sampling of the control samples.
    Only [Fe/H] bins with the wide binaries more than 5 are calculated. 
    \label{thin_r}} 
\end{figure*}

\begin{figure*}[htbp] 
    \centering %使图片居中显示
    \includegraphics[width=5in]{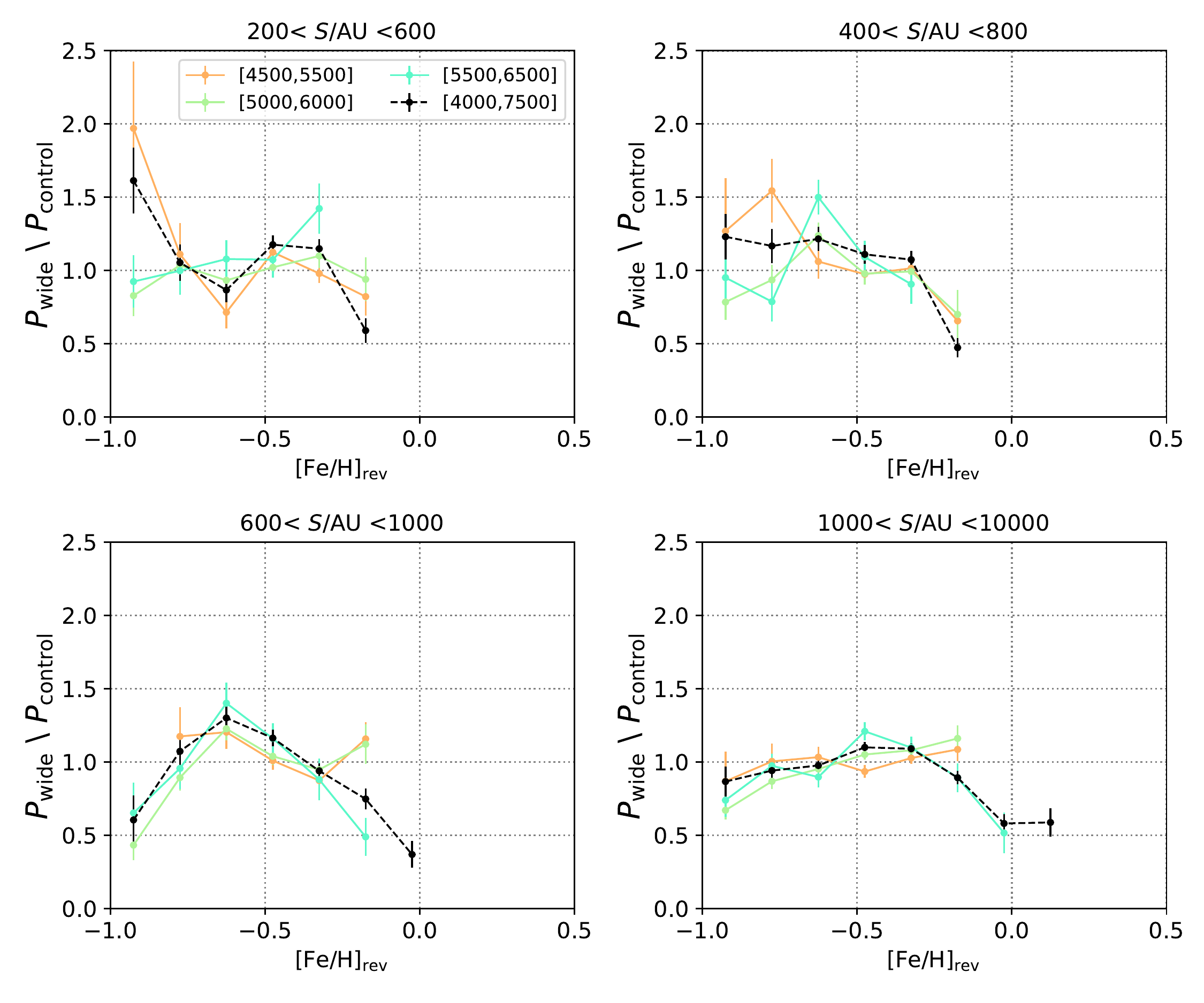}
    \caption{Same plots as Figure \ref{thin_r} but for the thick-disk samples. Notice that wide binaries hotter than 6500\,K as well as having separations less than 200\,AU are not contained in the thick-disk samples due to the small number.   
    \label{thick_r}} 
\end{figure*}

Figures \ref{thin_r} and \ref{thick_r} show the [Fe/H] dependence of the WBF for the thin and thick disks, respectively. 
Panels are ordered by the binary separations.
In each panel, colors represent the temperatures of primary. 
General results of 4000 -- 7500\,K are plotted in black.
[Fe/H] bins having less than 5 wide binaries are excluded. 

Most binaries in the first panel of Figure \ref{thin_r} are spectroscopically unresolved by LAMOST, considering its fiber size (3.3$^{\prime\prime}$). They are similar to the sample used in \citet{niu_bf} in terms of separations. 
Although not in a linear way, the WBF decreases with the increasing [Fe/H], which agrees with previous works (e.g., \citealp{raghavan2010}, \citealp{niu_bf}). 
Simultaneously, the impacts of the stellar temperatures gradually appear. 
The [Fe/H] dependence of the WBF keeps anti-correlated for the cold subsamples.
With the increasing temperature, the WBF becomes an ``arcuate'' shape with [Fe/H] for intermediate temperatures and then positively correlated with [Fe/H] for the hot subsamples.
This temperature-dependent gradient is particularly strong at [Fe/H] $>$ 0.  
It is reasonable to transform temperatures into stellar masses since the impacts of the surface abundance and gravity are negligible compared to those of the temperature. 
We stress that our results focus on the tendency of the WBF, not the absolute value.
The general tendency of WBF with 4000 -- 7000\,K is also plotted in black, as varying with the increasing separation. But it is quite dependent on the the proportions of different temperatures of the sample. In the thin-disk, we prefer to the mass-assigned results.

As for the thick-disk, the sample size is much smaller, particularly for wide binaries hotter than 6500\,K and having separations less than 200\,AU.
Unlike the dependence of the thin-disk samples, no impact of the temperature is found.
The WBF is roughly negative with [Fe/H] in the two top panels of Figure \ref{thick_r}. 
In the two bottom panels with larger separations, the WBF first increases and then decrease at [Fe/H] $\approx-0.5$, which agrees with \citet{widefb2021}.
The general tendency including the full temperature range is also plotted in black.

\section{Discussion}\label{secdis}

\subsection{Effect of the incompleteness}

The wide binary sample of this work comes from \citet{2021plx}. They set relative parallax uncertainties less than 20$\%$ and absolute parallax uncertainties less than 2 mas to guarantee the quality of astrometry. 
We also require the spectral signal-to-noise ratios larger than 20. 
Some faint and distant objects are not selected, which barely affects control samples but wide binaries. 
For primaries having the same apparent magnitude, distant binaries with small mass ratios ($q$) are more likely to be missed compared to nearby binaries.
The distance distributions may be different at different [Fe/H] bins, considering 
the vertical and radial metallicity gradient of the disk stars.

\begin{figure}[htbp] 
    \centering %使图片居中显示
    \includegraphics[width=3.5in]{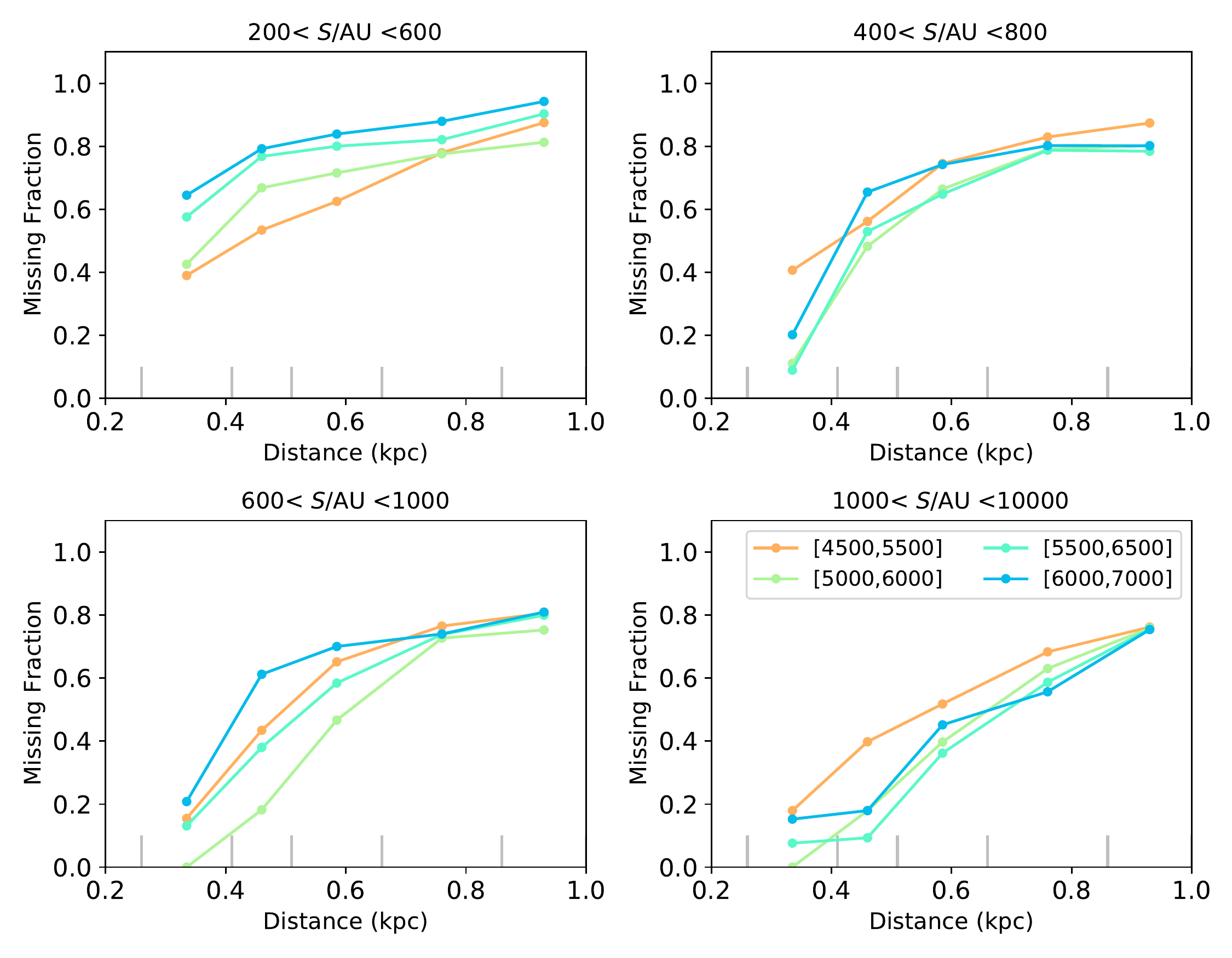}
    \caption{The missing fractions of wide binaries with faint secondaries as a function of distance. 
    The observed proportion of twin binaries is constant at certain ranges of mass and separation. 
    The missing fractions are calculated based on the proportion of twin binaries within 200\,pc. The higher missing fraction in the first panel is due to the target selection process of LAMOST. But it is independent of the stellar mass/temperature. 
    \label{incomp} }
\end{figure}

We use twin binaries to show that the mass-dependence of our results is not caused by this selection bias. 
For a resolved twin binary, its secondary of equal brightness should be observed once its primary is observed. 
The mass-ratio distribution is fixed at given ranges of mass and separation \citep{moe2017,2019twin}. 
Therefore, for each subsample in this work, we can roughly estimate the missing fraction of distant wide binaries using the observed proportion of twin binaries (with $G$ magnitude difference smaller than 0.4 mag) within 200\,pc. 
Notice that for wide binaries having separations less than 200\,AU, most of them are within 200\,pc and barely affected.
The missing fractions of various temperatures and separations as a function of distance are shown in Figure \ref{incomp}. 
In each panel, different temperature subsamples show a consistent trend.
Therefore, this incompleteness hardly affects results in this paper.
The reason for a higher missing fraction in the top left panel is due to the target selection process of LAMOST \citep{2015yuan_lasel}. 
For a given target star of $m$ magnitude, it is required to have no neighbor within 5 $^{\prime\prime}$ radius that is brighter than ($m$+1) mag. 
This probably makes the resolved binaries with small separations biased to small mass-ratio. But it is independent on the stellar mass.

\subsection{Comparison with previous work}\label{subseccomp}

\citet{2018dr2wide} and \citet{widefb2021} published their results about the metallicity dependence of the WBF based on \textit{Gaia} DR2. Although their samples and methods shared similarities to some extent, they came into different conclusions. 
\citet{widefb2021} attributed the discrepancies to their larger sample at larger distances and contributions of stellar ages. 
Our work shows that the disagreement is mostly caused by the different separation ranges.
As shown in Figure \ref{thin_r}, the general tendency varies with separation. 
Binaries of small separations ($s<$ 200\,AU) contributed an important part of \citet{2018dr2wide}, while \citet{widefb2021} 
focused on binaries of $10^3<s<10^4$\,AU.

Besides, we have found underestimations of [Fe/H] for low-temperature stars of 
the spectroscopic catalogs (LAMOST-LASP, APOGEE DR16, and GALAH DR3) used by 
\citet{2018dr2wide}, particularly for stars of T$_{\rm eff}<4500$\,K.
They used samples within 200\,pc, which owned a higher percentage of low-temperature binaries compared with the samples used in this work.
We perform measurements of the WBF before [Fe/H] corrections for the thin-disk sample.
In the left column of 4000 -- 5000\,K, the underestimation of [Fe/H] shifts the result towards the low-metallicity direction.
The impacts are very small in the middle and right columns, as expected. 

\begin{figure}[htbp] 
    \centering %使图片居中显示
    \includegraphics[width=3.5in]{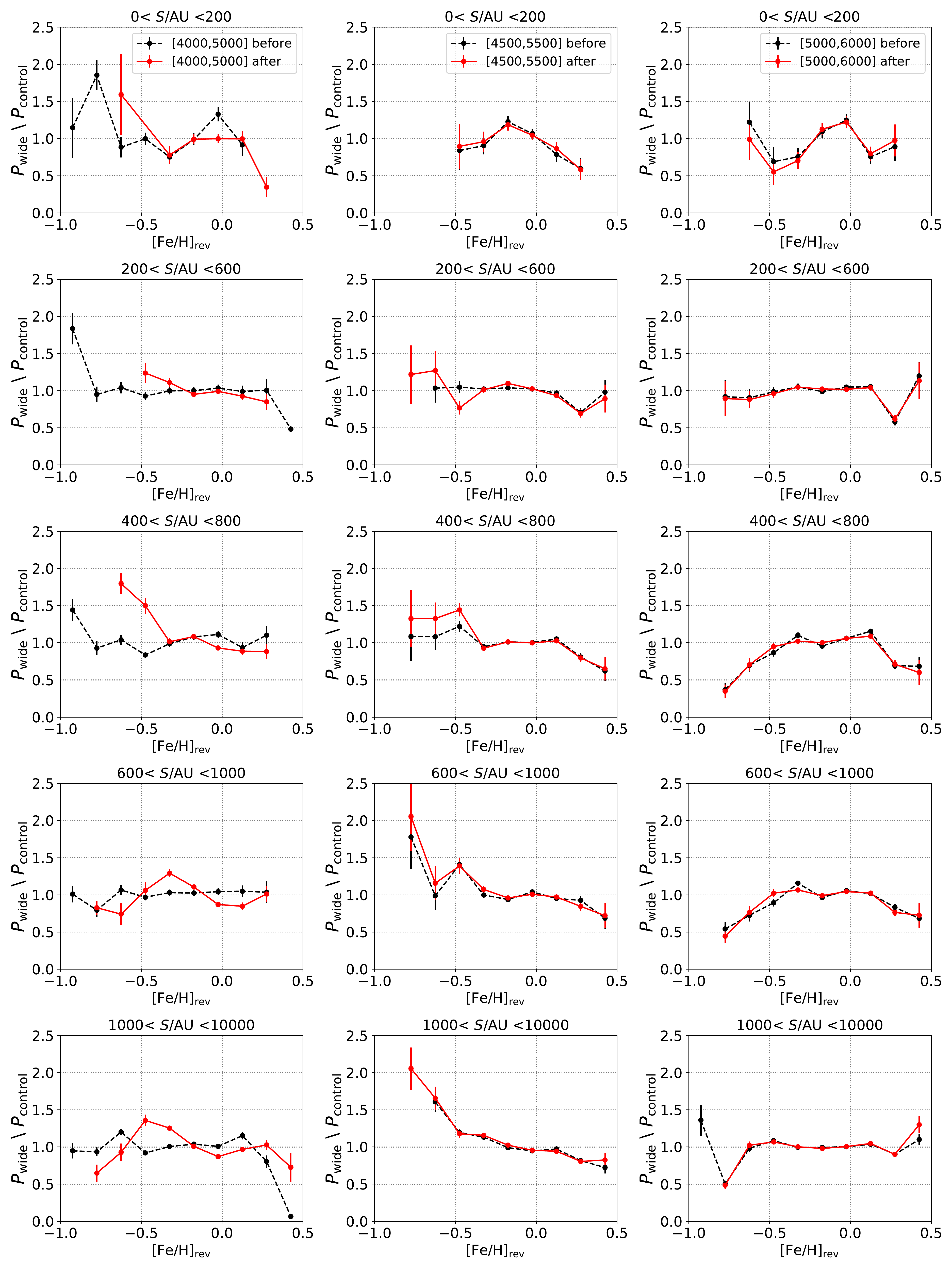}
    \caption{Results of the WBF tendency using [Fe/H] with and without the systematic corrections listed in Table \ref{tab1}.
    For left to right column, three temperature ranges are displayed. The separation range is labelled at the top of each panel. The impacts are very small for samples hotter than 4500\,K,  as expected.
    \label{4045} }
\end{figure}

\begin{figure*}[htbp] 
    \centering %使图片居中显示
    \includegraphics[width=6in]{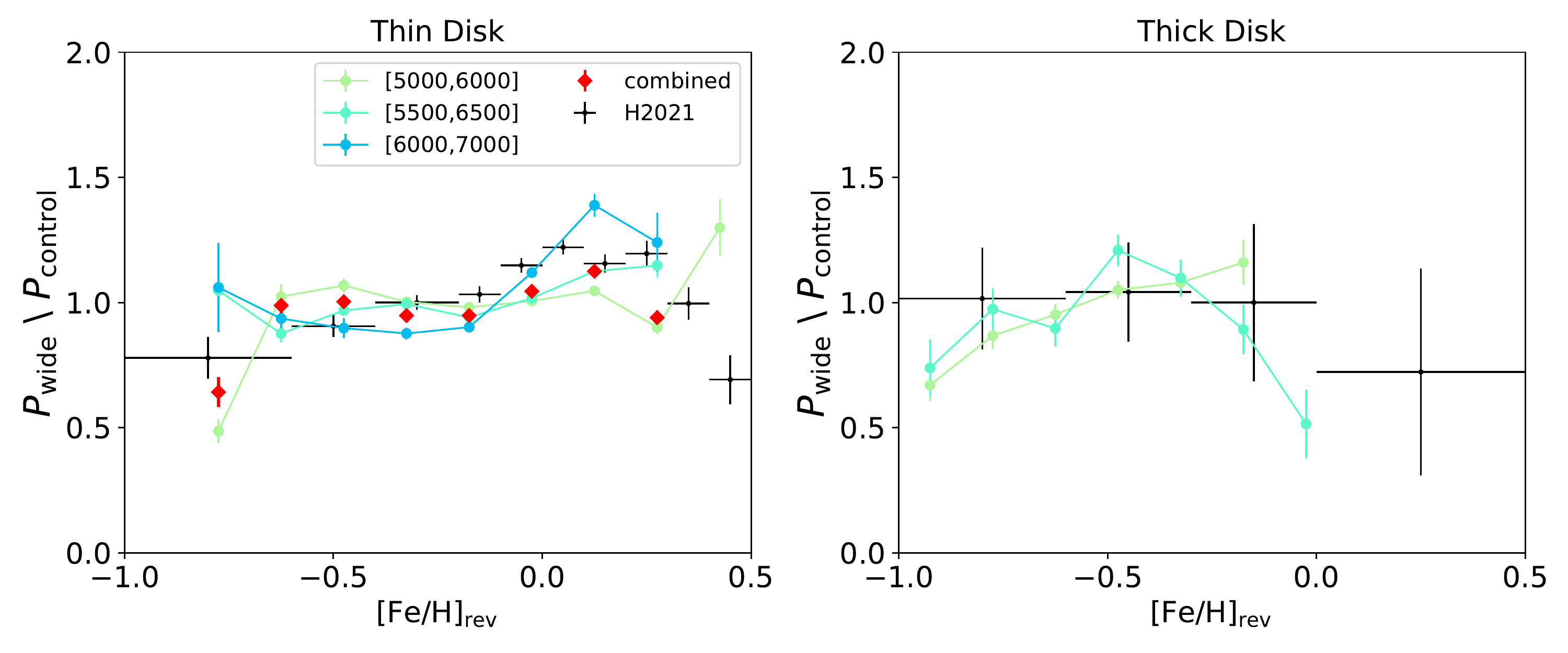}
    \caption{The [Fe/H] dependence of the WBF for stars between 5000 and 7000\,K. 
    Colors keep the same meaning as in Figure \ref{thin_r}. 
    Red points are combinations of 5000 -- 6000\,K and 6000 -- 7000\,K according to the temperature distributions of \citet{widefb2021}.
    In the thick disk, wide binaries hotter than 6500\,K are rare. 
    The black points are the normalized measurements from \citet{widefb2021},  where wide binaries from \textit{Gaia} DR2 and metallicities from LAMOST DR5 are applied. 
    \label{c_h} }
\end{figure*}

\citet{widefb2021} used metallicities from a single catalog and set temperatures between 5000 to 7000\,K, in which the systematic errors are relatively small.
We repeat their works and find that although \citet{widefb2021} did not specifically generate control samples following the same distance distribution of the wide binary samples, their distance distributions were quite similar by coincidence. 
We compare our results with theirs in Figure \ref{c_h} and 
find consistent results, after applying the same temperature range. 
Compared with \citet{widefb2021}, we complement the study by considering the impacts of temperature/mass. 
They also tested it with a narrower range of 5000 -- 6000\,K and found no big differences (see their APPENDIX A).
However, by using the wide binary catalog they provided, we find that their result of 6000 -- 7000\,K is close to ours.
\citet{widefb2021} and \citet{2022Hwang1} also studied the WBF inside different stellar populations and found the trends with metallicity are quite similar between the thin and thick disks.
% They stressed this conclusion in  by citing the results of \citet{widefb2021}.
After considering the effects from temperature, we can account for the differences between theirs and ours.

The mass-dependence of the metallicity dependence of the binary fraction has been hinted before. 
Using a binary sample with orbital periods $<$ 5 yr, \citet{2007Grether} found metal-poor stars own a higher binary fraction and further pointed out that this tendency is limited to relatively red stars ($B - V > 0.75$).
\citet{raghavan2010} reported that no relationship between metallicity and the possibility to have stellar companions exists for stars of $B - V < 0.625$.
\citet{2015Yuan} estimated the unresolved binary fraction within various ranges of stellar color and [Fe/H]. 
The redder subset shows a stronger anti-correlation between the binary fraction and metallicity 
than the bluer subset.
Our result shows that the metallicity dependence of the WBF changes with the temperature. In particular, it varies from the anti-correlation to positive correlation as the temperature increases from 4000\,K to 7500\,K.
Together with the above works, our result suggests that there are some physical processes behind the mass-dependence of the metallicity dependence of the binary fraction.
% This is supported by the above works considering that there were also some wide binaries in their samples.}

We agree with the possible explanations of the WBF tendency on metallicity suggested by \citet{widefb2021}.
We also remind that more factors should be involved when considering the mass-dependence.
For instance, high-mass stars are younger on average. The mass-dependence may be related to stellar ages.
The mass-dependence is not found in the thick-disk stars. 
We suspect that it is likely related to the different formation histories of the thin and thick disks.

\section{Summary}\label{summary}

Recently, there are disagreements about the metallicity dependence of the WBF.
By combining the \textit{Gaia} EDR3 wide binaries from \citep{2021plx} and LAMOST stellar parameters from \citet{xiangdd}, we present an independent study to investigate the metallicity dependence of the WBF.

We find that in addition to the metallicity and separation dependence of the WBF, mass-dependence exists in the thin disk, as demonstrated in Figure \ref{thin_r}.
For binaries with $s>200$\,AU, as the temperature rises from 4000\,K to 7500\,K, the relationship between the WBF and metallicity gradually changes from negative to positive. 
The general tendency agrees well with previous studies \citep{raghavan2010,niu_bf,widefb2021,moe2019anti}.
It shows anti-correlation at small separations, $s<200$\,AU for the thin-disk and $s<600$\,AU for the thick-disk. 
Then it becomes an ``arcuate'' shape at larger separations (hundreds to thousands of AU), peaking at [Fe/H] $\approx$ 0.1 for the thin-disk and [Fe/H] $\approx-0.5$ for the thick disk. 
Finally it becomes roughly flat for the thin-disk at $1000<s<10000$\,AU.
We notice the different results between the thin and thick disks, which may be related to 
their different formation histories.

The mass-dependence of the WBF barely gets noticeable until this work because previous sample sizes were not large enough to study the WBF in details under such high dimensions, such as [Fe/H], separation, and mass. 
The physical explanations still need further investigations.
This work provides new clues for theoretical works on the binary studies.

\begin{acknowledgments}
We acknowledge the referee for his/her valuable comments and suggestions that improved the 
quality of the paper significantly. 
We acknowledge helpful discussions with Dr. Kareem El-Badry, Prof. Maosheng Xiang, and Prof. Chao Liu.
We thank Zeiyang Pan and Zikun Lin for polishing up this paper. 

This work is supported by National Science Foundation of China (NSFC) under grant numbers 11988101, 11933004, and 12173007, National Key Research and Development Program of China (NKRDPC) under grant numbers 2019YFA0405504 and 2019YFA0405503, Strategic Priority Program of the Chinese Academy of Sciences under grant number XDB41000000.
We acknowledge the science research grants from the China Manned Space Project with NO. CMS-CSST-2021-A08 and CMS-CSST-2021-A09.

This work has made use of data products from the Guoshoujing Telescope (the Large Sky Area Multi-Object Fiber Spectroscopic Telescope, LAMOST). LAMOST is a National Major Scientific Project built by the Chinese Academy of Sciences. Funding for the project has been provided by the National Development and Reform Commission. LAMOST is operated and managed by the National Astronomical Observatories, Chinese Academy of Sciences. 

This work has made use of data from the European Space Agency (ESA) mission Gaia (https://www.cosmos.esa.int/gaia), processed by the Gaia Data Processing and Analysis Consortium (DPAC, https:// www.cosmos.esa.int/web/gaia/dpac/consortium). Funding for the DPAC has been provided by national institutions, in particular the institutions participating in the Gaia Multilateral Agreement.

This research made use of ASTROPY, a community-developed core PYTHON package for Astronomy \citep{astropy2013}.

\end{acknowledgments}

\appendix
\setcounter{table}{0}   %从0开始编号，显示出来表会A1开始编号
\setcounter{figure}{0}
% 定义编号格式，在数字序号前加字符“A"

\renewcommand\thefigure{\Alph{section}\arabic{figure}} 
\section{Verification of the disk populations}

\begin{figure*}[htbp] 
    \centering %使图片居中显示
    \includegraphics[width=7in]{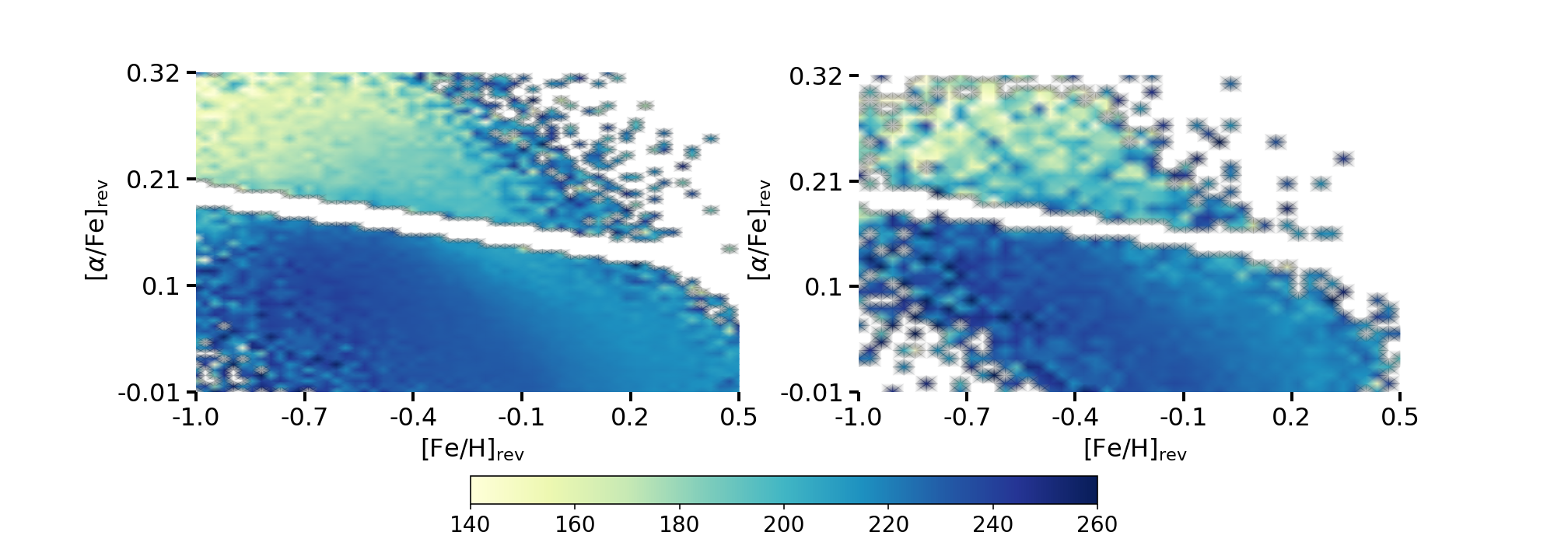}
    \caption{Distribution of the rotational velocity $V_{\rm \phi}$ in the [Fe/H]--[$\alpha$/Fe] plane.
    The left panel is the control sample, the right panel is the binary sample.
    Color indicates the $V_{\rm \phi}$. 
    The upper and lower colored areas are chemically defined thin and thick disks, respectively.
    Criteria of the thin and thick disks are listed in the main text.
    \label{kint}} 
\end{figure*}

We have used orbital rotational velocities ($V_{\rm \phi}$) to verify the efficacy of the criteria of disk populations. 
Distributions of the $V_{\rm \phi}$ in the [Fe/H]--[$\alpha$/Fe] plane are shown in Figure \ref{kint}.
Color bars donate the velocities, the darker the color, the higher the velocity. 
We find that the thick-disk stars have $V_{\rm \phi}$ values ranging from 140 km/s to 200 km/s, whereas those of the thin-disk stars are generally higher with a smaller dispersion and range from 220 km/s to 240 km/s.
The results are consistent with previous works (e.g. \citealp{cb2000,lee2011,jing2016,2020Anguiano}).

\bibliography{sample631}
\bibliographystyle{aasjournal}

\end{document}